\documentclass[3p,times]{elsarticle}

\usepackage{ecrc}

\usepackage{epstopdf}

\volume{00}

\firstpage{1}

\journalname{Nuclear Physics A}

\runauth{Author1 et al.}

\jid{nupha}

\jnltitlelogo{Nuclear Physics A}

\usepackage{graphicx}
\usepackage{amsmath,amssymb}
\begin{document}

\begin{frontmatter}

%% Title, authors and addresses

%% use the tnoteref command within \title for footnotes;
%% use the tnotetext command for the associated footnote;
%% use the fnref command within \author or \address for footnotes;
%% use the fntext command for the associated footnote;
%% use the corref command within \author for corresponding author footnotes;
%% use the cortext command for the associated footnote;
%% use the ead command for the email address,
%% and the form \ead[url] for the home page:
%%
%% \title{Title\tnoteref{label1}}
%% \tnotetext[label1]{}
%% \author{Name\corref{cor1}\fnref{label2}}
%% \ead{email address}
%% \ead[url]{home page}
%% \fntext[label2]{}
%% \cortext[cor1]{}
%% \address{Address\fnref{label3}}
%% \fntext[label3]{}

\title{Probing dense matter in compact star cores with radio pulsar data}

%% Single author (and collaboration) - please insert
\author{Mark G. Alford and Kai Schwenzer}
%\author[label1]{Mark Alford}
%\author[label1]{Kai Schwenzer}

%\fntext[col1] {A list of members of the XYZ Collaboration and acknowledgements can be found at the end of this issue.}
\address{Department of Physics, Washington University, St. Louis, Missouri, 63130, USA}

%% For multiple authors, replace the above by:

%\address[label1]{Department of Physics, Washington University, St. Louis, Missouri, 63130, USA}
%\address[label2]{Address 2}

\begin{abstract}
Astrophysical observations of compact stars provide, in addition to collider experiments, the other big source of information on matter under extreme conditions. The largest and most precise data set about neutron stars is the timing data of radio pulsars. We show how this unique data can be used to learn about the ultra-dense matter in the compact star interior. The method relies on astro-seismology based on special global oscillation modes (r-modes) that emit gravitational waves. They would prevent pulsars from spinning with their observed high frequencies, unless the damping of these modes, determined by the microscopic properties of matter, can prevent this. We show that for each form of matter there is a distinct region in a frequency/spindown-rate diagram where r-modes can be present. We find that stars containing ungapped quark matter are consistent with both the observed radio and x-ray data, whereas, even when taking into account the considerable uncertainties, neutron star models with standard viscous damping are inconsistent with both data sets and additional damping mechanisms would be required.
\end{abstract}

\begin{keyword}
%% keywords here, in the form: keyword \sep keyword
Dense matter \sep Compact stars \sep Pulsars 
%% MSC codes here, in the form: \MSC code \sep code
%% or \MSC[2008] code \sep code (2000 is the default)

\end{keyword}

\end{frontmatter}

%%
%% Start line numbering here if you want
%%
% \linenumbers

%% main text

\section{Introduction}
\label{intro}

Radio observations led to the discovery of pulsars
nearly fifty years ago \cite{Hewish:1968bj}. At the time the only explanation were neutron stars \cite{Lattimer:2004pg},
but since the formulation of QCD the study of strongly interacting matter under extreme conditions has become a very active research area. Whereas heavy ion experiments mainly probe the hot plasma phase, the interior of compact stars would be an ideal laboratory to study cold dense matter and many interesting options for the composition of such ultradense compact objects were proposed \cite{Alford:2007xm}, including most notably deconfined quark matter.
%\cite{Itoh:1970uw} ,Witten:1984rs}, where the elementary quark constituents are not confined in hadronic bound states. 
However, a search for novel phases of matter in compact stars requires appropriate astrophysical observables.
%Such novel phases
%could exist deep inside the core of the star where the
%density is high enough, and many possibilities have been
%proposed \cite{Alford:2007xm}. 
Since the initial discovery of pulsars a wealth
of radio pulsar data has been compiled \cite{Manchester:2004bp}
that is still our best source of information on compact stars and
even among the most precise data sets in physics. In particular, millisecond
pulsars, that can be billions of years old, are extremely stable systems
for which not only the frequency but also the spindown rate is known.
To use this timing data to learn about the star's interior composition poses a challenge, since the radio emission originates far away in the pulsar's magnetosphere, whereas no electromagnetic radiation can leave the dense core of a
compact star. Based on our recent article \cite{Alford:2013pma}, we show here how pulsar timing data can nevertheless
provide information about the dense matter in the interior and could allow us to discriminate different star compositions.
%However, this data could so far not be used to learn about the star's
%interior composition, since the radio emission originates far away
%in the magnetosphere. Here we show that pulsar timing data can indeed
%provide information about the dense matter in the interior and could
%help to discriminate different star compositions.

The method is based on astroseismology, which exploits that  mechanical oscillation modes can probe the opaque dense interior of the star. Even though such seismic oscillations cannot be observed, certain global modes emit
gravitational waves. R-modes \cite{Andersson:1997xt} are of particular
interest since they are unstable due to the Friedman-Schutz mechanism
\cite{Friedman:1978hf} and grow spontaneously. The gravitational waves emitted by r-mode oscillations may be directly detectable \cite{Alford:2012yn,Alford:2014pxa} in future interferometers \cite{Harry:2010zz}, but they also carry away angular
momentum and would prevent the spin-up of a star by accretion to millisecond frequencies if r-modes are not efficiently damped. 
Therefore, the fact that we observe millisecond pulsars means that such damping is present and the timing data of pulsars contains information about the damping in the interior. % of the star.
%In general, 
This damping stems from microscopic reactions within a given phase,
described particularly by its viscosities. It depends therefore not only on the equation of state, but on the low energy degrees of freedom, which can differ drastically for different forms of matter, and this provides an efficient way to discriminate them. 

\section{Dynamic versus static r-mode astroseismology}
\begin{figure}
%\begin{center}
\includegraphics*[width=8.cm]{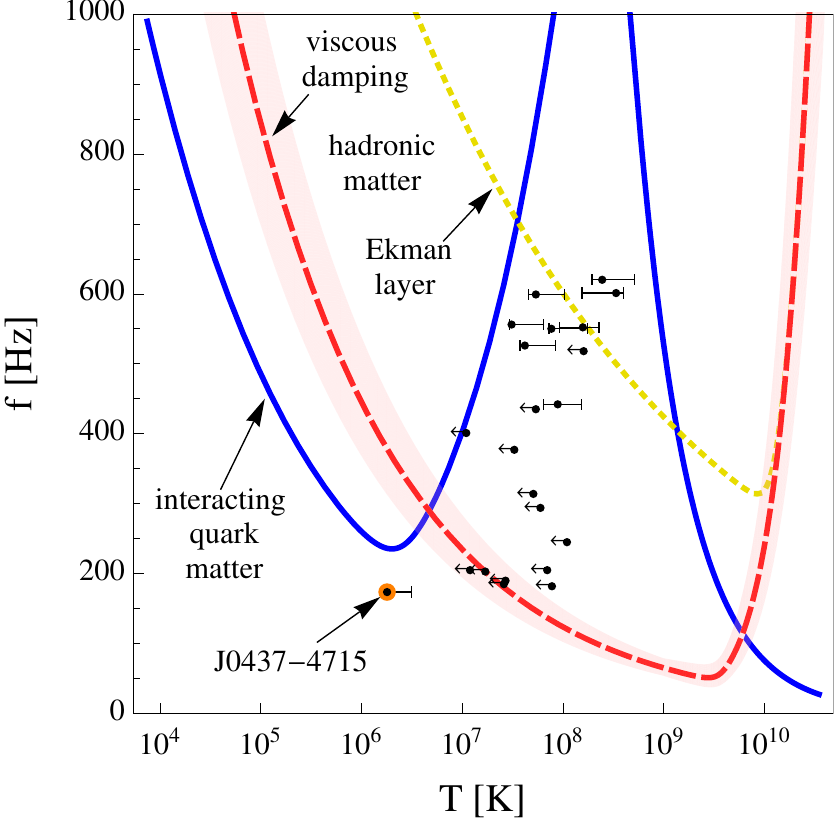}
\includegraphics*[width=8.cm]{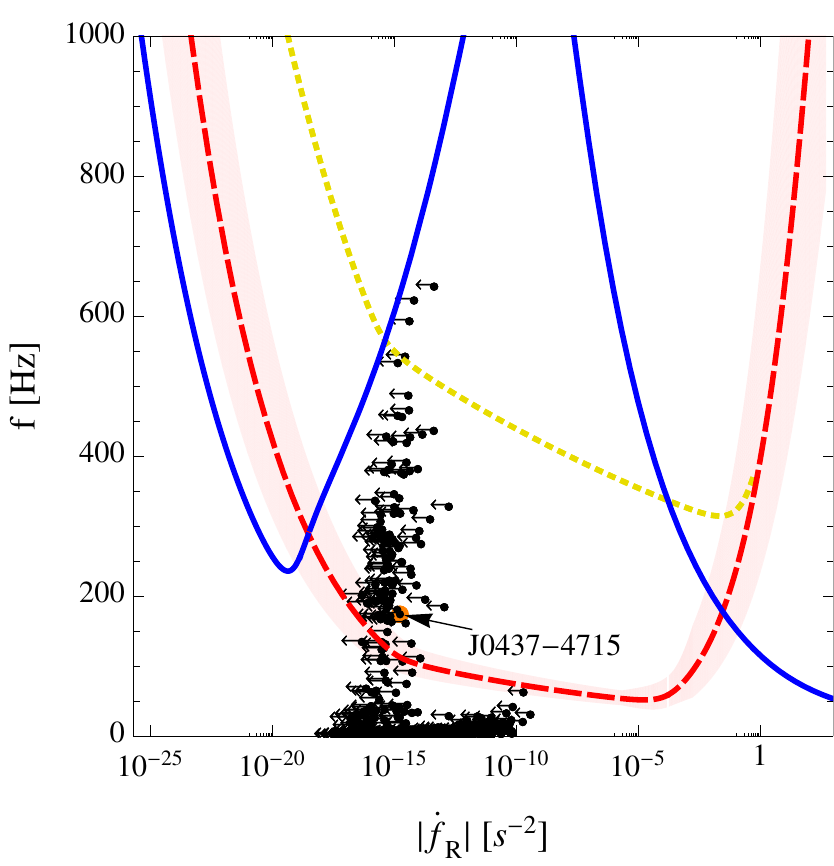}
\caption{
Boundaries of the r-mode instability
regions for different star compositions compared to pulsar data. \emph{Left:}
Standard static instability boundary compared to x-ray data \cite{Haskell:2012} %,Tomsick:2004pf}
with error estimates from different envelope models. %\cite{Gudmundsson:1983ApJ,Potekhin:1997}.
\emph{Right:} Dynamic instability boundary in timing parameter space
compared to radio data \cite{Manchester:2004bp} (all data points
are upper limits for the r-mode component of the spindown). The curves
represent: $1.4\, M_{\odot}$ neutron star (NS) with standard viscous
damping \cite{Shternin:2008es,Sawyer:1989dp} (dashed) and with additional
boundary layer rubbing \cite{Lindblom:2000gu} at a rigid crust (dotted)
as well as $1.4\, M_{\odot}$ strange star (SS) with long-ranged NFL
interactions causing enhanced damping \cite{Heiselberg:1993cr,Schwenzer:2012ga}
(using $\alpha_{s}\!=\!1$) (solid). More massive stars are not qualitatively
different, as is seen by the shaded band which shows for the standard neutron star an estimate of all uncertainties in both the micro- and macroscopic description. The encircled
points denote the only millisecond radio pulsar J0437-4715
with a temperature estimate.}
\label{fig:instability-regions}
%\end{center}
\end{figure}

We will first review the conventional way to perform r-mode seismology \cite{Andersson:2000mf}. Despite the damping due to viscosities or other dissipative mechanisms, r-modes are generally unstable at sufficiently large frequency in characteristic {\em static} instability regions in a $T$-$f$-diagram. 
%(shear viscosity generically becomes strong at low and bulk viscosity in the vicinity of a generally high resonance-%temperature). 
These are shown for a few models of dense matter on the left panel of figure \ref{fig:instability-regions}. These curves can be compared to data from low mass x-ray binaries (LMXB) \cite{Haskell:2012}
%,Tomsick:2004pf} 
which currently are spun up and heated by accretion and therefore emit x-ray radiation that allows to estimate their temperature. The data set is rather small and the temperatures are still uncertain since they require modeling the x-ray spectrum. %(the error bars only include the uncertainty due to two representative crust decompositions and should underestimate the actual uncertainty).
The dashed curve is the instability boundary for a neutron star with standard viscous damping. Most sources are clearly inside of this boundary and r-modes are therefore unstable at small r-mode amplitude $\alpha$. In this case the r-mode grows exponentially until it is stopped by an %non-linear, 
amplitude-dependent increase of the damping that saturates the amplitude at a finite value $\alpha_{\rm sat}$ \cite{Owen:1998xg,Bondarescu:2013xwa}. The only way to make this model compatible with the data would be that the corresponding saturation amplitude is so small that it hardly affects the spin evolution (the \emph{tiny r-mode} scenario). This would require a very low r-mode amplitude of the order $\alpha_{\rm sat} \lesssim 10^{-8}-10^{-7}$ \cite{Alford:2013pma,Mahmoodifar:2013quw}. However, currently proposed mechanisms for the saturation of r-modes can at most saturate r-modes at $\alpha_{\rm sat} \gtrsim 10^{-6}$ \cite{Bondarescu:2013xwa} so that the data cannot be explained by standard viscous damping in hadronic matter. Since neutron stars are structurally and compositionally complicated, and involve e.g. a crystalline crust, %superfluidity or large magnetic fields, these might lead to 
additional enhanced damping is possible in hadronic matter. Yet, in contrast to viscous damping often even the order of magnitude of these contributions is still uncertain and it will require further study to see if they are relevant. An example is the rubbing of the neutron fluid at the solid crust \cite{Lindblom:2000gu}, but as seen by the dotted curve, even under favorable assumptions \cite{Alford:2013pma} this mechanism cannot damp the instability in the fastest sources. In contrast many novel forms of matter feature enhanced viscous damping leading to a significantly reduced instability region (the \emph{no r-mode} scenario), as shown by the solid curve for a model of ungapped interacting quark matter \cite{Schwenzer:2012ga}.
%In general there are two alternative scenarios to explain the pulsar data: the
%damping is so strong that r-modes are completely damped (\emph{no
%r-mode} scenario), 

To probe dense matter in compact star cores with the distinct radio pulsar timing data set requires us to understand the spin evolution of millisecond pulsars in the presence of r-modes after the accretion stopped. The evolution is obtained from global conservation constraints \cite{Owen:1998xg} and includes the pulsar spin-frequency $f$, the dimensionless r-mode amplitude $\alpha$ \cite{Andersson:1997xt}, and, since the damping of r-modes heats the star, also of the temperature $T$ of the star \cite{Owen:1998xg}. The relevant observables such as the moment of inertia, the dissipative damping or the radiative cooling of the star depend on these evolution parameters via simple power laws. Therefore, an "effective theory" approach \cite{Alford:2012yn,Alford:2013pma} can be employed in order to describe the evolution of a pulsar,  which exploits that pulsars are effectively point sources and the complete information about the interior of the star can be encoded in a few ("low energy") constants, that are integrals over the entire star. Using this formalism one can derive general analytic expressions for the r-mode evolution \cite{Alford:2012yn,Alford:2013pma}.  
This is possible since the three coupled evolution parameters $f$, $\alpha$ and $T$ evolve on very different time scales. As discussed, if the r-mode is unstable, its amplitude saturates quickly 
%The amplitude initially grows exponentially and is then quickly stopped by a non-linear saturation mechanism which fixes it 
at a value $\alpha_{\rm sat}(f,T)$. It has been shown that for an arbitrary star composition the thermal evolution is always faster than the spin evolution \cite{Alford:2012yn}. Therefore, once the accretion stops, the star rapidly cools until it either leaves the instability region or until it reaches the thermal steady state curve, where heating is balanced by radiative cooling, along which it then very slowly spins down. The steady state depends on the r-mode saturation amplitude since the dissipative heating is large for a high amplitude mode. Following the evolution along this spindown curve until it leaves the static instability region allows to determine the r-mode spindown rate $\dot f_{\rm R}$ at this point and correspondingly to compute an analogous {\em dynamic}  instability region in a $\dot f_{\rm R}$-$f$ diagram. The striking feature of the dynamic boundary curves is that---like the static versions---they are completely independent of the saturation mechanism and the saturation amplitude \cite{Alford:2013pma}.

Dynamic instability regions for the star compositions discussed previously are shown on the right panel of fig. \ref{fig:instability-regions}. As can be seen these curves are qualitatively similar to the corresponding static instability regions on the left panel.
These curves can now be compared to the comprehensive timing data of millisecond radio pulsars \cite{Manchester:2004bp}, which had in the past been spun up in a LMXB system. It is important to stress that all data points are only upper limits since other spindown mechanisms should be present in these sources as well.
For the standard neutron star model (dashed) these upper bounds are again clearly inside the instability region. As discussed in \cite{Alford:2013pma}, the remarkable aspect of our method is that the mere upper spindown rate limits are sufficient to 
%use the data to 
constrain the interior composition. The reason for this is that millisecond pulsars are spun up at rather high temperature, as seen on the left panel of fig. \ref{fig:instability-regions}, and cannot cool out of the static instability region for all relevant saturation amplitudes $\alpha_{\rm sat} \gtrsim 10^{-10}$ \cite{Alford:2013pma}. This value is many orders of magnitude smaller than what  known saturation mechanisms can provide \cite{Bondarescu:2013xwa}. Consequently a source that happens to lie in the dynamic instability region must also lie inside of the static  region. If, in contrast, a source is outside of the dynamic instability region for a given star composition, r-modes cannot be the dominant spindown mechanism and such an instability region is therefore consistent with the data, as is the case for ungapped quark matter \cite{Schwenzer:2012ga} (solid).

In order to clearly discriminate different forms of dense matter requires control over the significant uncertainties in both our understanding of strongly interacting dense matter and of the properties of compact stars. A very favorable feature of the semi-analytic results for the instability regions \cite{Alford:2013pma} is that they can be extremely insensitive \cite{Alford:2010fd,Alford:2012yn} to the tremendous uncertainties in the underlying microscopic and macroscopic parameters (e.g. the bulk viscosity of dense matter involves strong interaction corrections and allows only an order of magnitude estimate). This insensitivity is illustrated for the instability regions of the minimal neutron star model in fig. \ref{fig:instability-regions}, where the uncertainty range is shown by the shaded bands and we use the uncertainty ranges for the underlying parameters estimated in \cite{Alford:2012yn}. These bands show that the minimal hadronic reference model is indeed inconsistent with the data. The uncertainties in the quark matter model are significantly bigger \cite{Schwenzer:2012ga}, but over large parts of the parameter space they are consistent with the data, so that within uncertainties ungapped quark matter would be a viable candidate for the interior of compact stars.

Another important requirement to use r-mode seismology to detect the enhanced damping of exotic forms of matter, is that we can rule out the tiny r-mode scenario, i.e. ensure r-modes are not saturated at such low amplitudes that they would be virtually irrelevant. For this it is interesting that even very low amplitude r-modes, that would not affect the spindown evolution, would significantly heat a source and the corresponding temperature is directly given in terms of $f$ and $\dot f_{\rm R}$ and independent of the r-mode saturation mechanism \cite{Alford:2013pma}. The corresponding upper bounds from the timing data lead to temperatures that are considerably larger than standard cooling models predict and which are large enough that it could be possible to detect the thermal x-ray radiation \cite{Alford:2013pma}. Therefore, either the detection or strict upper limits on the thermal x-ray observation of nearby isolated radio pulsars should tell us if r-modes can be present.

\section{Conclusions}

We have discussed how the comprehensive and precise timing data of radio pulsars can be used to probe the dense matter in the interior of compact stars. We show that standard neutron matter with mere viscous damping is inconsistent with the data and enhanced damping is required. As demonstrated, interacting ungapped quark matter is consistent with the data, but there are many other options, like in particular enhanced damping mechanisms in hadronic matter e.g. due to hyperons \cite{Lindblom:2001hd}, superfluidity, the neutron star crust \cite{Lindblom:2000gu} or large magnetic fields \cite{Rezzolla:1999he}. We will have to strongly improve our understanding of the damping in compact stars by detailed future studies, but with the combination of thermal x-ray, radio timing and future gravitational wave data we have an excellent data base, so that a clear discrimination of different forms of matter becomes a viable possibility.

\section*{Acknowledgements}
This material is based upon work supported by the U.S. Department of
Energy, Office of Science, Office of Nuclear Physics and High Energy
Physics under Award Number
\#DE-FG02-91ER40628  % Wash U high energy theory
and \#DE-FG02-05ER41375.

%% The Appendices part is started with the command \appendix;
%% appendix sections are then done as normal sections
%% \appendix

%% \section{}
%% \label{}

%% References
%%
%% Following citation commands can be used in the body text:
%% Usage of \cite is as follows:
%%   \cite{key}         ==>>  [#]
%%   \cite[chap. 2]{key} ==>> [#, chap. 2]
%%

%% References with BibTeX database:

\bibliographystyle{elsarticle-num}
%\bibliography{cs}

%% Authors are advised to use a BibTeX database file for their reference list.
%% The provided style file elsarticle-num.bst formats references in the required Procedia style

%% For references without a BibTeX database:

%\begin{thebibliography}{00}

%% \bibitem must have the following form:
%%   \bibitem{key}...
%%

%\bibitem{ref1} J. van der Geer, J.A.J. Hanraads, R.A. Lupton, J. Sci. Commun. 163 (2000) 51Ð59. 
%\bibitem{ref2} W. Strunk Jr., E.B. White, The Elements of Style, third ed., Macmillan, New York, 1979. 

%\end{thebibliography}

\end{document}